\documentclass[journal=nalefd,manuscript=letter,layout=preprint]{achemso}
\usepackage[version=3]{mhchem} 
\usepackage{graphicx}

\author{Yann Tison}
\altaffiliation{New address: Universit\'{e} de Pau et des Pays de l'Adour, IPREM - ECP CNRS UMR 5254, H\'{e}lioparc Pau-Pyr\'{e}n\'{e}es, 2 av. du Pr\'{e}sident Angot, 64053 Pau Cedex 9, France}
\affiliation{Laboratoire Mat\'{e}riaux et Ph\'{e}nom\`{e}nes Quantiques, Universit\'{e} Paris Diderot-Paris 7, Sorbonne Paris Cit\'{e}, CNRS, UMR 7162, \\ 10, rue A. Domon et L. Duquet, 75205 Paris 13, France}
\author{J\'{e}r\^{o}me Lagoute}
\email{jerome.lagoute@univ-paris-diderot.fr}
\author{Vincent Repain}
\author{Cyril Chacon}
\author{Yann Girard}
\affiliation{Laboratoire Mat\'{e}riaux et Ph\'{e}nom\`{e}nes Quantiques, Universit\'{e} Paris Diderot-Paris 7, Sorbonne Paris Cit\'{e}, CNRS, UMR 7162, \\ 10, rue A. Domon et L. Duquet, 75205 Paris 13, France}
\author{Fr\'{e}d\'{e}ric Joucken}
\author{Robert Sporken}
\affiliation{Research Center in Physics of Matter and Radiation (PMR), Universit\'{e} de Namur, 61 Rue de Bruxelles, 5000 Namur, Belgium}
\author{Fernando Gargiulo}
\author{Oleg V. Yazyev}
\affiliation{Institute of Theoretical Physics, Ecole Polytechnique F\'{e}d\'{e}rale de Lausanne (EPFL),
CH-1015 Lausanne, Switzerland}
\author{Sylvie Rousset}
\affiliation{Laboratoire Mat\'{e}riaux et Ph\'{e}nom\`{e}nes Quantiques, Universit\'{e} Paris Diderot-Paris 7, Sorbonne Paris Cit\'{e}, CNRS, UMR 7162, \\ 10, rue A. Domon et L. Duquet, 75205 Paris 13, France}

\title[An \textsf{achemso} demo]
{Grain Boundaries in Graphene on SiC(000$\mbox{\boldmath$\bar{1}$}$) Substrate}


\begin{document}

\begin{abstract}
Grain boundaries in epitaxial graphene on the SiC(000$\bar{1}$) substrate are studied using scanning tunneling microscopy and spectroscopy. All investigated small-angle grain boundaries show pronounced out-of-plane buckling induced by the strain fields of constituent dislocations. The ensemble of observations allows 
to determine the critical misorientation angle of buckling transition 
$\theta_c = 19~\pm~2^\circ$. Periodic structures are found among the flat large-angle grain boundaries. In particular, the observed $\theta = 33\pm2^\circ$ highly ordered grain boundary is assigned to the previously proposed lowest formation energy structural motif composed of a continuous chain of edge-sharing alternating pentagons and heptagons. This periodic grain boundary defect is predicted to exhibit strong valley filtering of charge carriers thus promising the practical realization of all-electric valleytronic devices. \\
\end{abstract}

\maketitle

The unique two-dimensional (2D) structure of graphene endows it with exceptional physical properties and potential applications \cite{Novoselov2004, Neto2009}. Large-area graphene sheets required by technology can be obtained by several scalable production routes such as epitaxial growth on the SiC substrate \cite{Berger2006} and chemical vapor deposition (CVD) on metal surfaces \cite{Li2009}. 
The obtained large-area samples are polycrystalline at micrometer length scales. Grain boundaries (GBs) intrinsically present in polycrystalline graphene significantly affect its properties\cite{Yazyev14} including electronic conductivity \cite{Yu2011,Tsen2012}, thermal conductivity \cite{Vlassiouk2011}, mechanical strength \cite{Grantab2010} and chemical reactivity \cite{Malola2010}. Understanding the structure and properties of GBs in graphene is crucial for the development of graphene-based nanoelectronics. 

Importantly, the atomic structure of GB defects depends 
strongly on the details of the synthetic method used for producing graphene.
In CVD graphene, transmission electron microscopy experiments have revealed the disordered aperiodic structure of GBs \cite{Huang2011,Kim2011} and recently ordered segments exhibiting van Hove Singularities have been observed \cite{Ma2014}.
In contrast, graphene grown on the SiC(000$\bar{1}$) substrate shows a strong tendency towards forming periodic GBs \cite{Varchon2008,Biedermann2009}, akin to those found in highly
oriented pyrolytic graphite \cite{Marchon1987, Albrecht1988, Simonis2002, vCervenka2009}. Investigating ordered one-dimensional defects in graphene is important as 
such structures have been proposed as a means of controlling 
electronic transport in graphene, thus realizing novel devices and extending the capabilities of traditional electronics \cite{Yazyev2010a,Gunlycke2011}.
Moreover, such structures can be used as a testbed for investigating
fundamental theoretical predictions regarding 2D
materials. One such phenomenon is the buckling transition resulting from the interplay between in-plane stress fields, produced by dislocations constituting the GB defects, and the out-of-plane deformation \cite{Carraro93,Yazyev2010,Liu2010}.  

In this Letter, we report a comprehensive scanning tunneling microscopy (STM) investigation of GBs in epitaxial graphene on SiC(000$\bar{1}$) combining atomic resolution imaging and local spectroscopy. A broad statistics of the observed GBs allowed determining the critical misorientation angle for the transition between flat and buckled regimes. A scanning tunneling spectroscopy (STS) investigation of a highly regular large-angle GB reveals the presence of localized electronic states. First-principles calculations allowed this GB to be assigned to the previously predicted lowest formation energy structural motif 
composed of a continuous chain of edge-sharing alternating pentagons and heptagons. Non-equilibrium Green's function calculations show that charge carriers transmitted across this periodic defect acquire high valley polarization, thus making it a prospective component of 
valleytronic devices.

\begin{figure}[]
\includegraphics[width=.6\columnwidth]{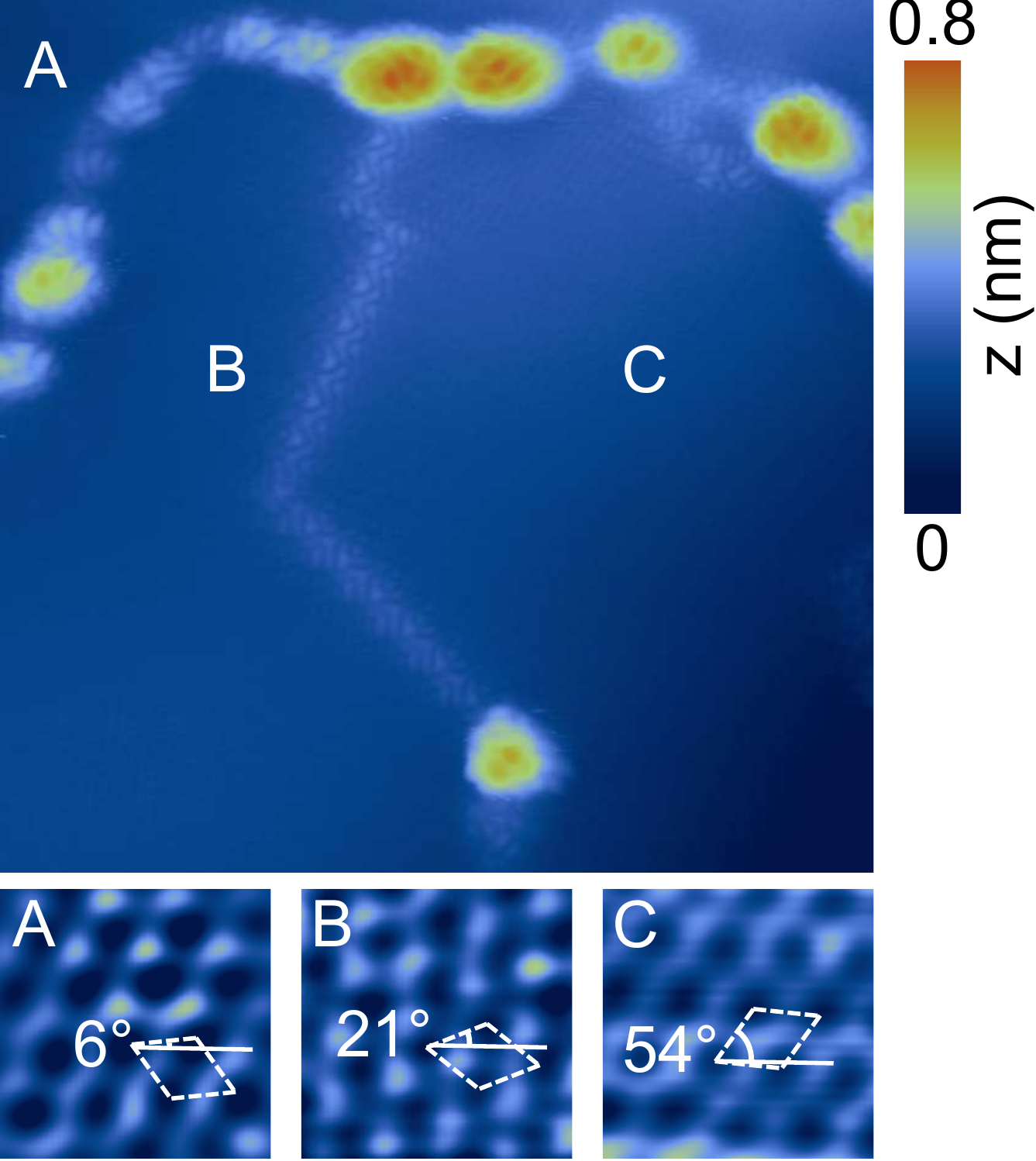}
\caption{Raw STM image of epitaxial graphene grown on SiC(000$\bar{1}$) substrate (15$\times$15~nm$^2$, $U = 0.75$~V, $I = 500$~pA) that exhibit three single-crystalline domains with different lattice orientations (labeled A, B and C).  The magnified atomic-resolution images (1$\times$1 nm$^2$) show the orientation of the three domains with respect to a common reference vector. Buckled (between domains A and B, and domains A and C) and flat (between domains B and C) GBs are observed.}
\label{Fig1}
\end{figure}

%\section*{EXPERIMENT}
The sample was obtained by annealing a SiC(000$\bar{1}$) substrate in UHV at 1300$^{\circ}$C under a Si flux as previously reported \cite{Bommel1975, Varchon2008}. This method produces multilayer graphene samples with a rotational disorder between adjacent layers. The number of layers was larger than five, as verified by Auger spectroscopy.
STM/STS measurements were performed using an Omicron Nanotechnology low temperature STM operating under UHV conditions (less than $10^{-10}$ mbar) at liquid nitrogen temperature. Local $dI/dV$ spectra were recorded using a lock-in amplifier with a modulation at 711~Hz and 30~mV amplitude. The sample was introduced into the UHV system and outgassed at a temperature of 880$^{\circ}$~C to remove residual adsorbed molecules. All measurements were performed with electrochemically etched tungsten tips.

%\section*{RESULTS AND DISCUSSION}
Figure~\ref{Fig1} shows a representative topographic STM image of epitaxial graphene on SiC(000$\bar{1}$) covering a 15$\times$15~nm$^2$ area. The image reveals three single-crystalline domains with different lattice orientations
separated by three GB defects forming a junction in the upper part of the imaged area. Each GB is characterized by a misorientation angle $\theta = \theta_1 - \theta_2$, where $\theta_1$ and $\theta_2$ are the lattice orientations of the two domains with respect to a common reference vector. 
Below, we adhere to the convention of Ref.~\cite{Yazyev2010} defining misorientation angle in the range $0^{\circ}<\theta<60^{\circ}$ (small-angle grain boundaries thus correspond to $\theta$ close to $0^{\circ}$ or $60^{\circ}$).
We note that the observed GBs show some striking differences in  appearance. 
The GB separating domains B and C is characterized by a misorientation angle $\theta=33\pm2^{\circ}$ and exhibits a highly ordered structure with clear atomic-scale periodicity.
The two linear segments ca.~5~nm long are separated by $120^\circ$ turn and show identical STM topographies. This periodic structure is characterized by a typical height of 0.07~nm at tunneling bias $U = 0.75$~V. 
In contrast, the two other GBs separating A and B ($\theta=16\pm3^{\circ}$) and A and C ($\theta=48\pm4^{\circ}$) exhibit much larger topographic contrasts with typical heights of 0.26~nm and 0.36~nm, respectively, measured at $U = 0.75$~V. These defects appear as chains of ``humps'' in which periodicity is respected only approximately. The observed significant topographic contrast is assigned to the out-of-plane deformation. Such buckling in the third dimension represents a physical mechanism unique to 2D systems, which allows relieving the in-plane elastic strain produced by topological defects \cite{Seung88,Carraro93}. The grain-boundary buckling provides a mechanism of relieving the in-plane elastic strain created by individual dislocations that constitute one-dimensional GB defects. The dominant long-range part of this strain field depends on the magnitude of the Burgers vector of dislocations, rather than on the atomic structure of their cores. In large-angle GBs ($\theta \sim 30^\circ$), the distance between neighboring dislocations is comparable to their Burgers vectors, which results in effective mutual cancellation of the strain fields produced by dislocations. Such GBs structures do not undergo buckling. However, for larger and smaller values of GB misorientation angles $\theta$, the distance between neighboring dislocations in GB structures increases,\cite{Yazyev2010} giving rise to buckled structures beyond certain values of critical misorientation angles $\theta_c$.

Corrugation induced by well-separated dislocations have been observed in both epitaxial \cite{Coraux08} and suspended graphene \cite{Lehtinen13}. In the former case, individual dislocations appear in STM image as hump-like features similar to the ones observed in our work.   
    
\begin{figure}[]
\centering
\includegraphics[width=0.6\columnwidth]{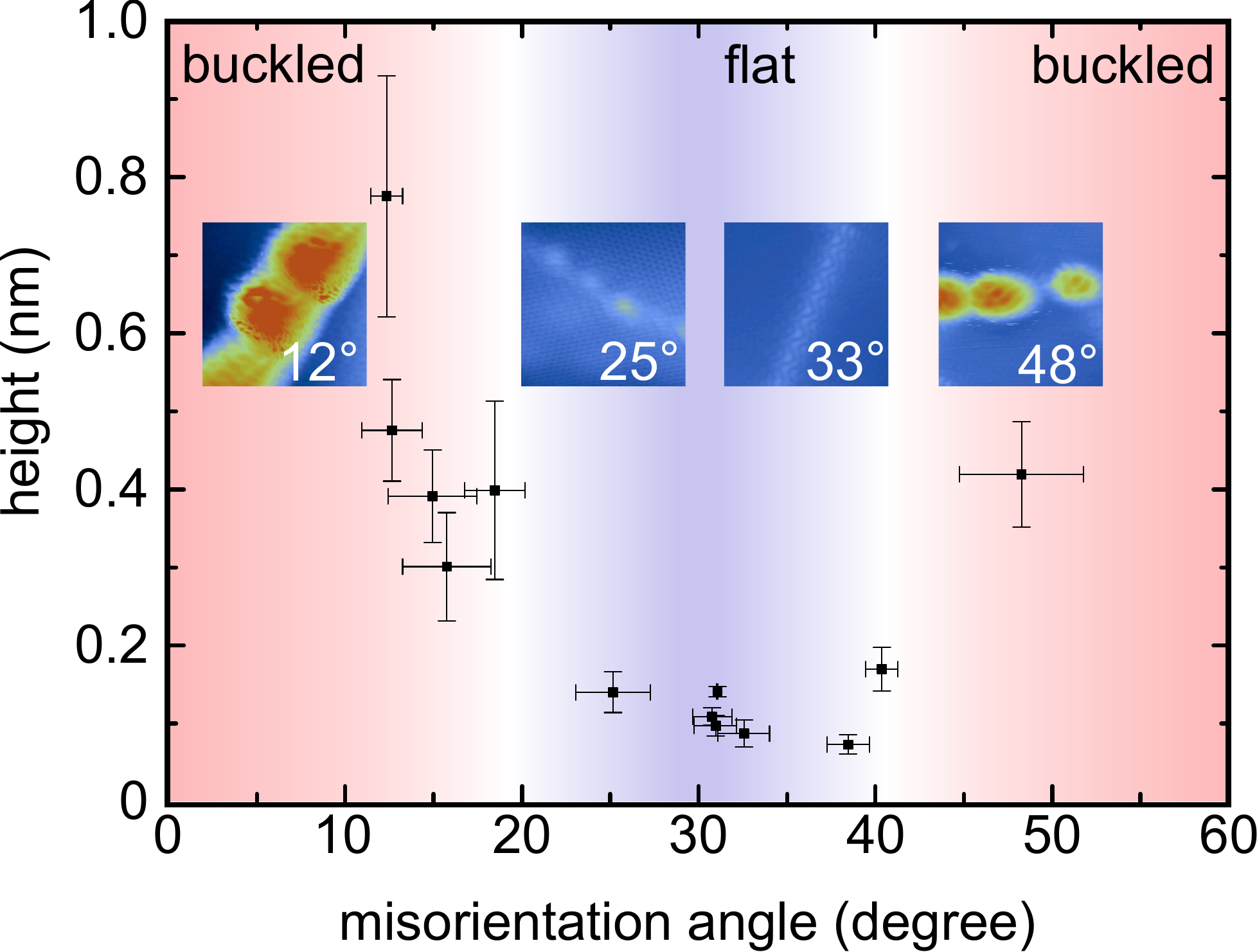}
\caption{Apparent height of grain boundaries (measured in STM images at $U = 1$~V, $I = 500$~pA) as a function of misorientation angle $\theta$. The insets show representative STM images (5$\times$5~nm$^2$) of buckled and flat GBs corresponding to different misorientation angles $\theta$ (same color scale as in Fig.~\ref{Fig1}).}
\label{Fig2}
\end{figure}

The total number of investigated GBs is 14, thus allowing us to determine the critical misorientation angle at which transition 
between flat and buckled regimes takes place. Figure~\ref{Fig2} reports 
the results of our measurements presented as apparent height of GBs
as a function of misorientation angle $\theta$. Since STM probes 
both the surface geometry and electronic effects with the latter being 
very sensitive to bias voltage, all measurements have
been performed at the same bias voltage $U = 1$~V.  The uncertainties in measuring
the angle are due to piezo drift, and the error bars on the height are due to variations of the apparent height of local maxima.
The results in Fig.~\ref{Fig2} clearly show that all observed GBs can be divided into two distinct categories defined by an apparent height smaller or larger than 0.2~nm. This feature does not depend critically on the applied bias voltage as (see supporting information Fig.~S1). As exemplified in the inset, well-defined periodic structures are observed for the large-angle GBs ($\theta \sim 30^\circ$). 
Their low apparent height allows assigning these structures to the flat regime. The particular case of $\theta = 25\pm2^{\circ}$ is more ambiguous as its atomic-scale structure is superimposed with a superstructure of 2.6~nm periodicity. In contrast, all small-angle GBs
($\theta \sim 10^\circ$ and $\theta \sim 50^\circ$) display pronounced 
``humps'' of much larger apparent height which allows assigning them to 
the buckled regime. From the ensemble of data presented we conclude that the transitions between the flat and buckled regimes occur within the 
ranges $18^\circ < \theta < 25^{\circ}$ and $40^\circ < \theta < 48^{\circ}$. We further assume that the critical values of the transition 
are symmetric with respect to $\theta = 30^{\circ}$. This assumption 
is based on the symmetry of graphene lattice and the fact that  
the dependence of GB energies on $\theta$ is roughly symmetric with 
respect to $\theta = 30^\circ$ \cite{Yazyev2010}. By combining the two ranges of buckling transition we estimated the value of the critical angle $\theta_c = 19\pm2^{\circ}$, with corresponding symmetric transition taking place at $\theta_c = 41\pm2^{\circ}$. The obtained value fully agrees with the results of first-principles calculations in which only the periodic GB models characterized by 
$\theta = 21.8^\circ$, $32.2^\circ$ and $38.2^\circ$ belong to the flat phase \cite{Yazyev2010}.

\begin{figure}[]
\centering
\includegraphics[width=.6\columnwidth]{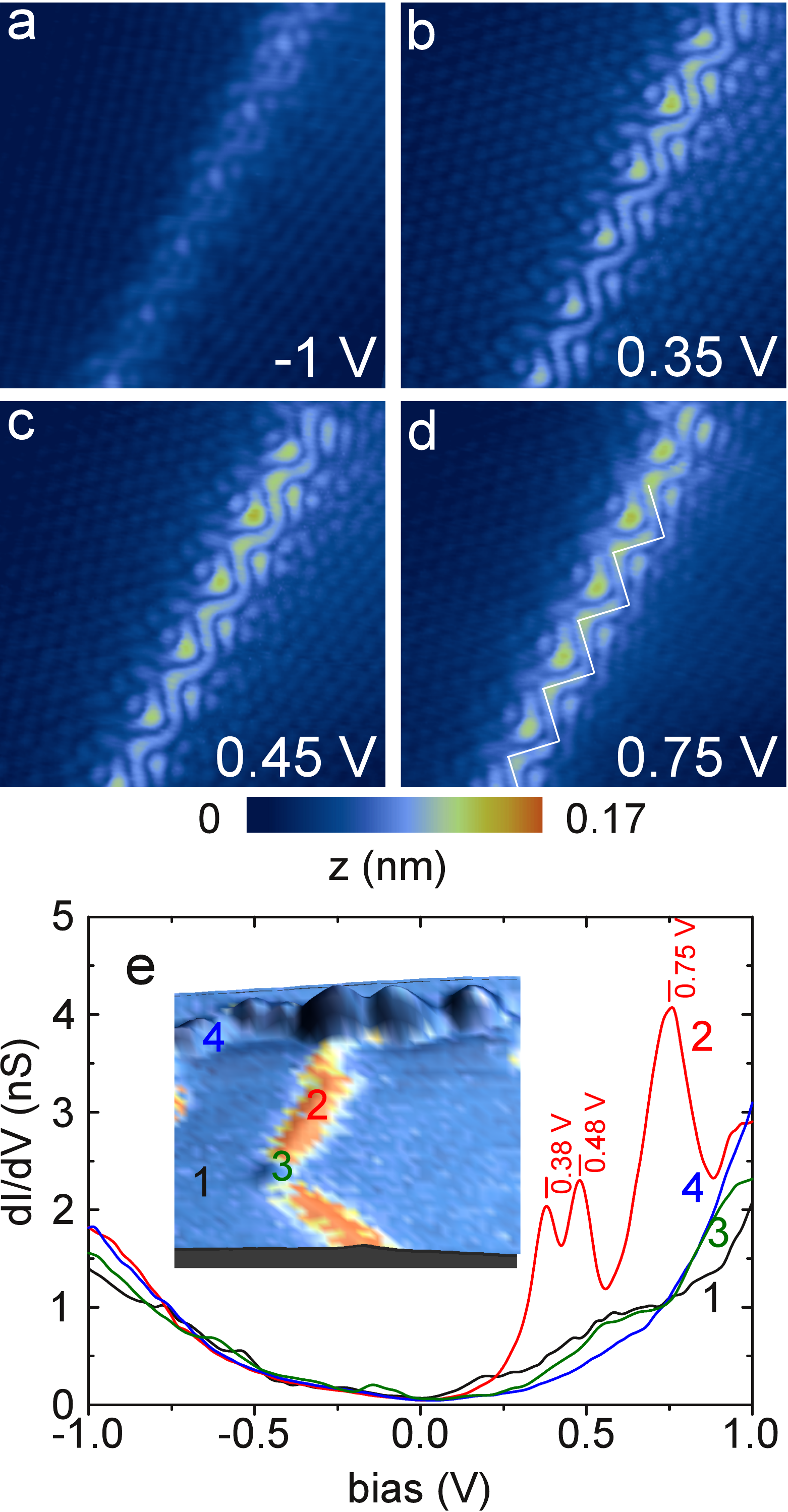}
\caption{(a--d) Topographic STM images (4.5$\times$4.5~nm$^2$) of the $\theta = 33\pm 2^{\circ}$ grain boundary shown in Fig.~\ref{Fig1} measured at different bias voltages $U$ and the tunneling current $I = 500$~pA. The white broken line in (d) is a guide to the eye indicating the zigzag pattern of the central part of the grain boundary. (e) $dI/dV$ spectra measured on the graphene sheet (black), on the grain boundary (red), at the kink of the grain boundary (green) and on the buckled grain boundary between domains A and C areas in Fig.~\ref{Fig1} (blue). Inset: Composite 3D topographic image (5$\times$5~nm$^2$) color coded with the conductance map at 0.75~V (blue: low conductance, red: high conductance) showing the area where the $dI/dV$ spectra were measured.}
\label{Fig3}
\end{figure}

We now focus our attention on the electronic properties of the highly ordered flat GB with $\theta=33\pm2^{\circ}$ discussed above.
Figure~\ref{Fig3}(a--d) shows the topographic images obtained at different bias voltages ($U = -1$~V, 0.35~V, 0.45~V and 0.75~V). At three positive bias voltages we observe the same periodic pattern, while a different motif is seen at negative bias. The bias dependence of the image suggests significant differences of the electronic structure at positive and negative biases. Figure~\ref{Fig3}(e) compares $dI/dV$ spectra measured above the grain boundary (red curve) and inside a single-crystalline domain of graphene far from the GB (black curve). Inside the graphene domain the $dI/dV$ spectrum has a typical V-shaped curve with a minimum close to the Fermi level corresponding to the Dirac point. In contrast, sharp and intense peaks are observed in the $dI/dV$ spectrum of the GB at 0.38~V, 0.48~V and 0.75~V. Interestingly, these peaks vanish at the $120^\circ$ junction of the GB [center left of Fig.~\ref{Fig1}] as shown by the green curve in Fig.~\ref{Fig3}(e). Similarly, no $dI/dV$ peak was observed in the case of buckled GB between domains A and B of Fig.~\ref{Fig1} [blue curve in Fig.~\ref{Fig3}(e)]. 
The three peaks measured in the $dI/dV$ spectrum are therefore characteristic of the $\theta=33\pm2^{\circ}$ GB. The spatial distribution of the electronic states was probed by mapping the $dI/dV$ signal that corresponds to the local density of states (LDOS). The inset of Fig.~\ref{Fig3}(e) shows such a conductance map recorded at 0.75~V and projected on the topography image of the GBs. The conductance map clearly shows that the peak at 0.75~V corresponds to an electronic state localized on the linear segments of the GB. 
Detailed information on the atomic scale variation of the experimental spectroscopy and $dI/dV$ maps are given in the supporting information Fig.~S2.

\begin{figure}[]
\centering
\includegraphics[width=\columnwidth]{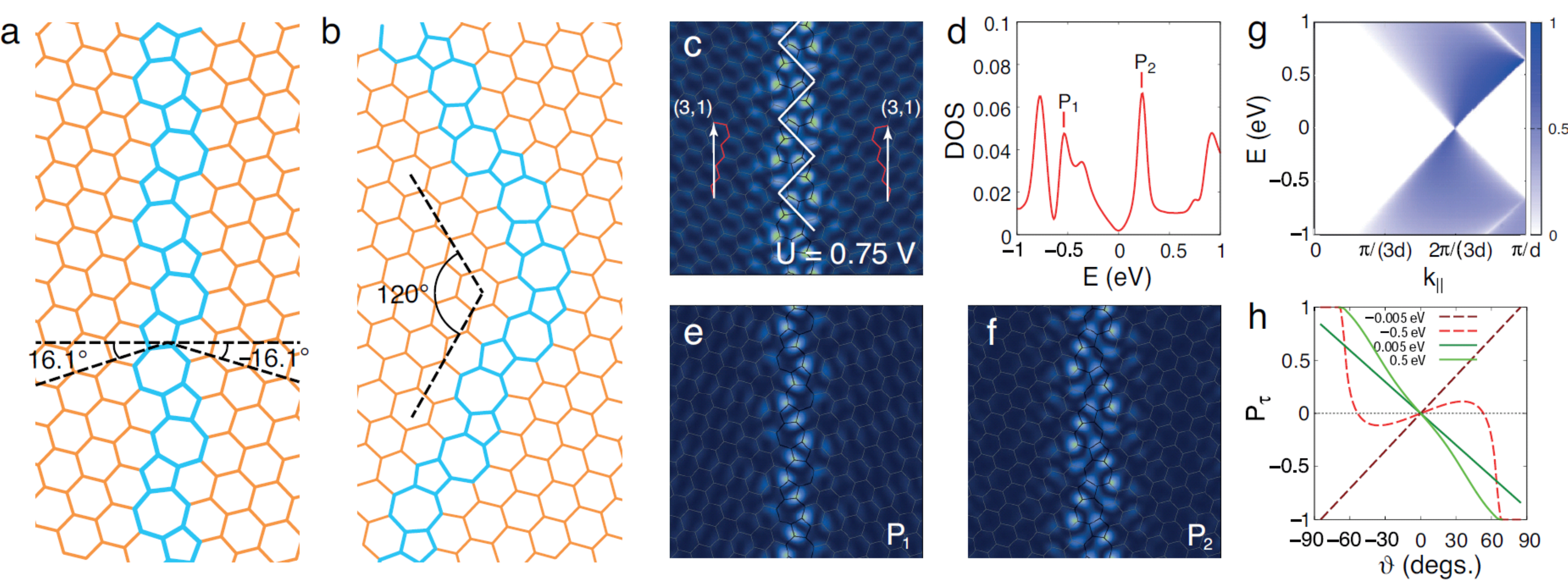}
\caption{Atomistic models of (a) the $\theta=32.2^{\circ}$ symmetric GB ($\theta_1 = -\theta_2 =16.1^{\circ}$) and (b) its $120^\circ$ junction. Pentagons and heptagons are highlighted. (c) Simulated STM image of the $\theta=32.2^{\circ}$ GB (5$\times$5~nm$^2$, bias voltage $U = 0.75$~V). The white broken line is a guide to the eye (same scale as in Fig.~\ref{Fig3}(d)) indicating the zigzag pattern of the central part of the grain boundary. Its periodicity is further indicated by matching vectors \cite{Yazyev2010a} (3,1) = 0.9 nm. (d) Density of states plot of the $\theta=32.2^{\circ}$ GB calculated from first principles. (e,f) Simulated $dI/dV$ maps at energies corresponding to the DOS peaks labeled P$_1$ and P$_2$ in panel (d). (g) Charge-carrier transmission probability $T(k_{||},E)$ across the $\theta=32.2^{\circ}$ GB as a function of momentum along the GB $k_{||}$ and energy $E$ calculated using the tight-binding model. Half of the Brillouin zone ($0 < k_{||} < \pi/d$, where is $d = 0.89$~nm is the periodicity of $\theta=32.2^{\circ}$ GB) and one valley are shown for clarity. (h) Valley polarization $P_\tau(\vartheta,E)$ of charge-carriers transmitted through the GB defect as a function incidence angle $\vartheta$ at different energies. }
\label{Fig4}
\end{figure}

The measured misorientation angle $\theta=33\pm2^{\circ}$ provides 
a strong suggestion of the possible structure of this GB. Indeed, a
symmetric GB structure with $\theta=32.2^{\circ}$ ($\theta_1 = -\theta_2 =16.1^{\circ}$) proposed in Ref.~\cite{Yazyev2010} represents
a continuous chain of edge-sharing pentagons and heptagons [Fig.~\ref{Fig4}(a)]. Such a structure has been recently observed by STM on graphene on SiO$_{2}$ \cite{Ma2014}. The $dI/dV$ spectra reported on SiO$_{2}$ differ from the spectra we measured on SiC. We attribute this to the difference of sample roughness and structure in proximity to the GB and possibly the difference of supporting substrate. All carbon atoms in this structure maintain their original
threefold coordination. Importantly, the observed $120^\circ$ turn can
be introduced with minimal changes to the arrangement of pentagons and heptagons [Fig.~\ref{Fig4}(b)]. In order to verify this hypothesis we
perform first-principles simulations \cite{SIESTA} of the STM images within the Tersoff-Hamann approximation \cite{Tersoff85}. The simulated STM image,
assuming bias voltage $U = 0.75$~V, is shown in Fig.~\ref{Fig4}(c). The qualitative features appear to be identical to those of the 
experimental STM images recorded at positive bias voltages [Fig.~\ref{Fig3}(b--d)]. In particular, experimental images reveal a periodic zigzag pattern with bright spots in the corners. The simulated STM images show extra nodal lines crossing the zigzag, which are likely not resolved in the experimental images. However, we note the periodicity of the zigzag patterns (0.9 nm) in actual and simulated STM images coincides, as indicated by the broken lines displayed in Fig.~\ref{Fig3}(d) and Fig.~\ref{Fig4}(c) that have the same scale. This STM signature is due to the localized electronic states with large weights on the carbon atoms belonging to 
pentagons and heptagons. The properties of these localized electronic 
states are further elaborated with the help of the density-of-states (DOS) plot shown in Fig.~\ref{Fig4}(d). In particular, we observe a pair of van Hove singularities at positive and negative energies [labeled P$_1$ at $E=-0.55$~eV and P$_2$ at $E=0.25$~eV in Fig.~\ref{Fig4}(d)], which is characteristic to GBs in graphene \cite{Gargiulo14}. We note, however, that the position of the peaks do not agree with the $dI/dV$ spectrum shown in Fig.~\ref{Fig3}(e). This disagreement is likely to be a consequence of compressed energy scales typical to ordinary density-functional-theory calculations as well as due to tip-induced effects \cite{McEllistrem93}.
The combination of these two factors shifts both peaks observed in STS measurements to higher absolute energies compared to our first-principles
calculations. The experimental counterpart of the negative-energy peak P$_1$ is likely to be out of scale in the measurements. The simulated $dI/dV$ maps in Fig.~\ref{Fig4}(e,f) nevertheless show qualitative agreement with corresponding experimental images [Fig.~\ref{Fig3}(a--d) and supporting information Fig.~S2]. 

Misorientation angle is defined by the orientation
of graphene seeds, which is related mostly to the kinetic aspects of the initial stages of the growth process. On the contrary, the atomic structure of GBs can be strongly influenced by local equilibrium aspects.
The observation of highly ordered GBs implies that the growth process of our epitaxial graphene on SiC(000$\bar{1}$) allows approaching local thermodynamic equilibrium closer than, for instance, CVD growth on the Cu surface \cite{Li2009}, which typically results in more disordered large-angle GBs \cite{Huang2011,Kim2011}. Indeed, the morphology of the polycrystalline graphene and the structure of GBs present in it depend strongly on the growth method used to synthesize graphene \cite{Biro2013}. Moreover, the observation of the $\theta=32.2^{\circ}$ symmetric GB
is not surprising considering the fact that this structural motif,
first predicted in Ref.~\cite{Yazyev2010}, corresponds to the minimum of the GB energy vs. misorientation angle curve
($E_f = 0.284$~eV/\AA). 

Ordered line defects in graphene have been predicted to
exhibit novel transport phenomena originating from the conservation of momentum $k_{||}$ along the defect \cite{Yazyev2010a,Gunlycke2011}.
Below, we reveal the transport properties of
the $\theta=32.2^{\circ}$ GB by means of a theoretical investigation.
Periodicity of the discussed $\theta=32.2^{\circ}$ GB is defined by matching vector indices $(n_{\rm L},m_{\rm L}) = (n_{\rm R},m_{\rm R}) = (3,1)$, which identifies this line defect structure as class Ib GB, according to the classification introduced in Ref.~\cite{Yazyev2010a}. No transport gap is expected in this case, but the two valleys $\tau = +1$ and $\tau = -1$  are separated in $k_{||}$ suggesting that this GB structure can exhibit valley polarized transport properties similar to the 5--5--8 line defect discussed previously \cite{Lahiri2010,Gunlycke2011}. Figure~\ref{Fig4}(g) shows the ballistic charge-carrier transmission probability $T(k_{||},E)$ across this GB as a 
function of momentum $k_{||}$ and energy $E$ calculated using the non-equilibrium Green's function approach and the nearest-neighbor tight-binding Hamiltonian $H = - t \sum_{\langle i,j \rangle} [ c_{i}^\dagger c_{j} + {\rm h.c.} ]$ ($t = 2.7$~eV) \cite{Neto2009}.
The magnitude of the transmission probability $T(k_{||},E)$ 
shows significant variations upon changing $k_{||}$, and hence 
the incidence angle $\vartheta$ of ballistic charge carriers. This 
effect is particularly pronounced for electrons ($E > 0$) where $T(k_{||},E)$ increases monotonically with increasing $k_{||}$ at constant $E$. The opposite tendency is observed for the hole charge carriers ($E < 0$) with the dependence practically vanishing at 
$E \approx -0.5$~eV. The calculated valley polarization of ballistic charge carriers expressed as a function of their incidence angle $\vartheta$
$$P_\tau(\vartheta,E) = \frac{T_{\tau=+1}(\vartheta,E)-T_{\tau=-1}(\vartheta,E)}{T_{\tau=+1}(\vartheta,E)+T_{\tau=-1}(\vartheta,E)}$$
for several different values of energy $E$ is shown in Fig.~\ref{Fig4}(h). Apart from the case of high-energy hole charge carriers ($E=-0.5$~eV), $P_\tau$ exhibits an almost linear dependence on $\vartheta$ with complete valley polarization achievable at oblique incidence angles. Such a simple dependence is largely due to the 
absence of strong resonant backscattering characteristic to the 5--5--8 line defect \cite{Chen2014}.  

In conclusion, our combined STM and theory study reveals that grain-boundary defects in epitaxial graphene on the SiC(000$\bar{1}$) substrate tend to form highly ordered structures. This opens novel opportunities for 
tailoring electronic and transport properties of graphene, in particular, for realizing valleytronic devices \cite{Rycerz2007,Xiao2007} in which valley polarization is controlled by all-electric means. In addition, the observation of a sufficiently large number of flat and corrugated grain boundaries allowed determining the critical misorientation angle of buckling transition 
$\theta_c = 19~\pm~2^\circ$, thus confirming previous theoretical predictions.

\begin{acknowledgement}
V.R. thanks the Institut Universitaire de France for support. Y. T. thanks the Labex SEAM program No.~ANR-11-LABX-086 for financial support in the
framework of the Program No.~ANR-11-IDEX-0005-02.
F.G. and O.V.Y. acknowledge financial support of the Swiss National Science Foundation (grant No.~PP002P\_133552). First-principles calculations have been performed at the Swiss National Supercomputing Centre (CSCS) under projects s443 and s515. 
\end{acknowledgement}

\begin{suppinfo}
Additional experimental data for bias dependent imaging of the GBs and $dI/dV$ maps of the $\theta=33\pm2^{\circ}$ GB.
\end{suppinfo}

%\bibliography{GrainBoundaries}

\providecommand{\latin}[1]{#1}
\providecommand*\mcitethebibliography{\thebibliography}
\csname @ifundefined\endcsname{endmcitethebibliography}
  {\let\endmcitethebibliography\endthebibliography}{}

\begin{tocentry}
\includegraphics{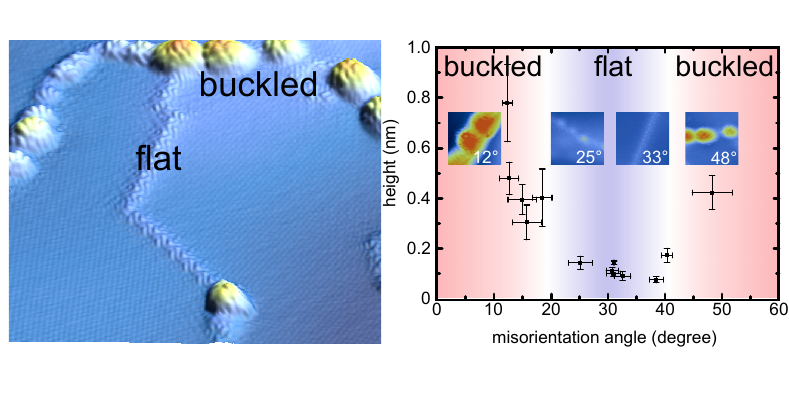}
\end{tocentry}

\end{document}